\documentclass{cpbtex}
\usepackage{authblk}

\usepackage{float}
\usepackage{graphicx}
\usepackage{hyperref}
\usepackage{xcolor}
\usepackage{lineno}

\usepackage{xcolor}

\newcommand{\FIG}[1] {Figure~\ref{#1}}

\newcommand{\AFF}[1]{$^{\foreach\d[count=\ni]in{#1}{\ifnum\ni=1\ref{\d}\else,\ref{\d}\fi}}$}

\def\cmpc{\mathrm{cm}^{-3}\, \mathrm{pc}}

\title{Atlas of dynamic spectra of fast radio burst FRB~20201124A}

\begin{document}

\definecolor{blue}{RGB}{51,87,255}

\author{Bo-Jun Wang\textsuperscript{1,2,3}, Heng Xu\textsuperscript{2}, \textcolor{black}{Jin-Chen Jiang}\textsuperscript{2}, \textcolor{black}{Jiang-Wei Xu}\textsuperscript{1,2,3}, \textcolor{black}{Jia-Rui Niu}\textsuperscript{2,4}, \textcolor{black}{Ping Chen}\textsuperscript{1,3,5}, \textcolor{black}{Ke-Jia Lee}\textsuperscript{1,2}\thanks{E-mail: kjlee@pku.edu.cn}, \textcolor{black}{Bing Zhang}\textsuperscript{6,7}\thanks{Email: bing.zhang@unlv.edu}, \textcolor{black}{Wei-Wei Zhu}\textsuperscript{2}\thanks{Email: zhuww@nao.cas.cn}, \textcolor{black}{Su-Bo Dong}\textsuperscript{1}, \textcolor{black}{Chun-Feng Zhang}\textsuperscript{2}, \textcolor{black}{Hai Fu}\textsuperscript{8}, \textcolor{black}{De-Jiang Zhou}\textsuperscript{2,4}, \textcolor{black}{Yong-Kun Zhang}\textsuperscript{2,4}, \textcolor{black}{Pei Wang}\textsuperscript{2}, \textcolor{black}{Yi Feng}\textsuperscript{2,9}, \textcolor{black}{Ye Li}\textsuperscript{10}, \textcolor{black}{Dong-Zi Li}\textsuperscript{11}, \textcolor{black}{Wen-Bin Lu}\textsuperscript{12}, \textcolor{black}{Yuan-Pei Yang}\textsuperscript{13}, R.~N.~Caballero\textsuperscript{1,3}, \textcolor{black}{Ce Cai}\textsuperscript{14}, \textcolor{black}{Mao-Zheng Chen}\textsuperscript{15}, \textcolor{black}{Zi-Gao Dai}\textsuperscript{16}, A.~Esamdin\textsuperscript{15}, \textcolor{black}{Heng-Qian Gan}\textsuperscript{2}, \textcolor{black}{Jin-Lin Han}\textsuperscript{2}, \textcolor{black}{Long-Fei Hao}\textsuperscript{17}, \textcolor{black}{Yu-Xiang Huang}\textsuperscript{17}, \textcolor{black}{Peng Jiang}\textsuperscript{2}, \textcolor{black}{Cheng-Kui Li}\textsuperscript{14}, \textcolor{black}{Di Li}\textsuperscript{2,18}, \textcolor{black}{Hui Li}\textsuperscript{2}, \textcolor{black}{Xin-Qiao Li}\textsuperscript{14}, \textcolor{black}{Zhi-Xuan Li}\textsuperscript{17}, \textcolor{black}{Zhi-Yong Liu}\textsuperscript{15}, \textcolor{black}{Rui Luo}\textsuperscript{19}, \textcolor{black}{Yun-Peng Men}\textsuperscript{20}, \textcolor{black}{Chen-Hui Niu}\textsuperscript{2}, \textcolor{black}{Wen-Xi Peng}\textsuperscript{14}, \textcolor{black}{Lei Qian}\textsuperscript{2}, \textcolor{black}{Li-Ming Song}\textsuperscript{14}, \textcolor{black}{Jing-Hai Sun}\textsuperscript{2}, \textcolor{black}{Fa-Yin Wang}\textsuperscript{21}, \textcolor{black}{Min Wang}\textsuperscript{17}, \textcolor{black}{Na Wang}\textsuperscript{15}, \textcolor{black}{Wei-Yang Wang}\textsuperscript{3}, \textcolor{black}{Xue-Feng Wu}\textsuperscript{10}, \textcolor{black}{Shuo Xiao}\textsuperscript{14}, \textcolor{black}{Shao-Lin Xiong}\textsuperscript{14}, \textcolor{black}{Yong-Hua Xu}\textsuperscript{17}, \textcolor{black}{Ren-Xin Xu}\textsuperscript{1,3,22}, \textcolor{black}{Jun Yang}\textsuperscript{21}, \textcolor{black}{Xuan Yang}\textsuperscript{10}, \textcolor{black}{Rui Yao}\textsuperscript{2}, \textcolor{black}{Qi-Bin Yi}\textsuperscript{14}, \textcolor{black}{You-Ling Yue}\textsuperscript{2}, \textcolor{black}{Dong-Jun Yu}\textsuperscript{2}, \textcolor{black}{Wen-Fei Yu}\textsuperscript{23}, \textcolor{black}{Jian-Ping Yuan}\textsuperscript{15}, \textcolor{black}{Bin-Bin Zhang}\textsuperscript{21,24}, \textcolor{black}{Song-Bo Zhang}\textsuperscript{10}, \textcolor{black}{Shuang-Nan Zhang}\textsuperscript{14}, \textcolor{black}{Yi Zhao}\textsuperscript{14}, \textcolor{black}{Wei-Kang Zheng}\textsuperscript{12}, \textcolor{black}{Yan Zhu}\textsuperscript{2}, \textcolor{black}{Jin-Hang Zou}\textsuperscript{21,25}}
\affil{\textsuperscript{1}{ \textcolor{black}{Kavli Institute for Astronomy and Astrophysics, 
Peking University, Beijing 100871, China}},\\
\textsuperscript{2}{ \textcolor{black}{National Astronomical Observatories, Chinese Academy of Sciences, Beijing 100101, China}},\\
\textsuperscript{3}{ \textcolor{black}{Department of Astronomy, Peking University, Beijing 100871, China}},\\
\textsuperscript{4}{ \textcolor{black}{University of Chinese Academy of Sciences, Chinese Academy of Sciences, Beijing 100049, China}},\\
\textsuperscript{5}{ Department of Particle Physics and Astrophysics, Weizmann Institute of Science, Rehovot 76100, Israel},\\
\textsuperscript{6}{ Nevada Center for Astrophysics, University of Nevada, Las Vegas, NV 89154, USA},\\
\textsuperscript{7}{ Department of Physics and Astronomy, University of Nevada, Las Vegas, NV 89154, USA},\\
\textsuperscript{8}{ Department of Physics \& Astronomy, University of Iowa, Iowa City, IA 52242, USA},\\
\textsuperscript{9}{ \textcolor{black}{Zhejiang Lab, Hangzhou, Zhejiang 311121, China}},\\
\textsuperscript{10}{ \textcolor{black}{Purple Mountain Observatory, Chinese Academy of Sciences, Nanjing 210008, China}},\\
\textsuperscript{11}{ TAPIR, Walter Burke Institute for Theoretical Physics, Mail Code 350-17, Caltech, Pasadena, CA 91125, USA},\\
\textsuperscript{12}{ Department of Astronomy, University of California at Berkeley, 
Berkeley, CA 94720, USA},\\
\textsuperscript{13}{ \textcolor{black}{South-Western Institute For Astronomy Research, Yunnan University, Yunnan 650504, China}},\\
\textsuperscript{14}{ \textcolor{black}{Key laboratory of Particle Astrophysics, Institute of High Energy Physics, Chinese Academy of Sciences, Beijing 100049, China}},\\
\textsuperscript{15}{ \textcolor{black}{Xinjiang Astronomical Observatory, Chinese Academy of Sciences, Urumqi 830011, China}},\\
\textsuperscript{16}{ \textcolor{black}{University of Science and Technology of China, Anhui 230026, China}},\\
\textsuperscript{17}{ \textcolor{black}{Yunnan Observatories, Chinese Academy of Sciences, Kunming 650216, China}},\\
\textsuperscript{18}{ \textcolor{black}{Guizhou Normal University, Guiyang 550001, China}},\\
\textsuperscript{19}{ CSIRO Space and Astronomy, Epping, NSW 1710, Australia},\\
\textsuperscript{20}{ Max-Planck institut f$\ddot{u}$r Radioastronomie, Auf Dem H$\ddot{u}$gel, Bonn, 53121, Germany},\\
\textsuperscript{21}{ \textcolor{black}{School of Astronomy and Space Science, Nanjing University, Nanjing 210093, China}},\\
\textsuperscript{22}{ \textcolor{black}{State Key Laboratory of Nuclear Physics and Technology, School of Physics, Peking University, \\Beijing 100871, China}},\\
\textsuperscript{23}{\textcolor{black}{Shanghai Astronomical Observatory, Chinese Academy of Sciences, Shanaghai 200030, China}},\\
\textsuperscript{24}{\textcolor{black}{Key Laboratory of Modern Astronomy and Astrophysics (Nanjing University), Ministry of Education, China}},\\
\textsuperscript{25}{ \textcolor{black}{College of Physics, Hebei Normal University, Shijiazhuang 050024, China}},\\

}

\renewcommand*{\Affilfont}{\small\it}

\maketitle
\begin{abstract}
Fast radio bursts (FRBs) are highly dispersed millisecond-duration radio bursts\cite{Lorimer07Sci,Zhang20Nature}, of which the physical origin is still not fully understood. FRB~20201124A is one of the most actively repeating FRBs. In this paper, we present the collection of 1863 burst dynamic spectra of FRB~20201124A measured with the Five-hundred-meter Aperture Spherical radio Telescope (FAST).  The current collection, taken from the observation during the FRB active phase from April to June 2021, is the largest burst sample detected in any FRB so far. The standard \textcolor{black}{PSRFITs} format is adopted, including dynamic spectra of the burst, and the time information of the dynamic spectra, in addition, mask files help readers to identify the pulse positions are also provided. The dataset is available in Science Data Bank, with the link {\url{https://www.doi.org/10.57760/sciencedb.j00113.00076}}.
\end{abstract}

\hspace{2em}\textbf{Keywords:} Fast radio burst, FAST


\section{Introduction}

Fast radio bursts (FRBs) are bright, millisecond duration pulses with dispersion measures (DM) mostly well in excess of Galactic values, since first discovered in 2007\cite{Lorimer07Sci}, more than 800 FRBs \textcolor{black}{have} been detected so far and 27 of them can emit repeating bursts\cite{Luo2020Natur,Niu2022Natur} {(\url{https://www.herta-experiment.org/frbstats/catalogue})}. Currently, 19 FRBs have been localized to host galaxies {(\url{https://frbhosts.org/})}. Although the physical mechanism of FRBs still \textcolor{black}{remains} unknown, FRB~200428\cite{CHIME_atel,CHIME2020,STARE2_CHIMEburst_atel,Bochenek2020} produced by Galactic magnetar SGR~J1935+2154 suggests that some of the FRBs can be emitted by magnetars\cite{2020Natur.587...63L,Zhang20Nature}. Among all the FRBs, FRB~20201124A, which \textcolor{black}{was} discovered by CHIME\cite{CHIME2021ATel}, \textcolor{black}{has} been frequently studied recently. Its radio bursts show rich pulse structures\cite{xuheng2021,2022MNRAS.509.2209M}. Through dynamic spectra, researchers investigated the scintillation time-scale of FRB~20201124A\cite{2022MNRAS.509.3172M}. Efforts had also been made to localize its host galaxy\cite{2021ATel14515....1D,2021ATel14516....1K,2021ATel14518....1X,2021ATel14538....1W}. \par

  Dynamic spectra record the FRB intensity as a function of time and frequency. Dynamic spectra contain information of FRB intrinsic emission properties as well as density fluctuation of interstellar and inter-galactic medium. We noted that there is lack of a systematic collection of dynamic \textcolor{black}{spectra} for FRBs. In this paper, we present the dynamic spectra data of FRB~20201124A which covers 1863 pulses detected by our team.

\section{Observation, data acquisition, and analysis}

We used \textcolor{black}{the} Five-hundred-meter Aperture Spherical radio Telescope (FAST)\cite{jiang2019} to monitor FRB~20201124A from April to June in 2021. The FAST 19-beam Pulsar back-end covers 1.0-1.5 GHz in frequency band and has $a$ system temperature about 20 to 25~K \cite{jiang2020}. The data were recorded using the digital back-end based on the Re-configurable Open Architecture Computing Hardware-2 (\textsc{Roach2}) board \cite{HA16} with temporal resolution of 49.152 $\mu$s or 196.608 $\mu$s and frequency resolutions of 122.07 kHz. \par

Our data processing contains two major steps, searching for single pulses and post processing to form the dynamic spectra. Firstly, we searched for the FRB candidates offline with software package \textsc{TransientX}. Frequency channels affected by radio frequency interference (RFI) were removed. The data were de-dispersed in the dispersion measure (DM) range of 380-440\,$\cmpc$ with a step of 0.1\,$\cmpc$ since FRB~20201124A is a known repeater. The pulse width is searched from 0.1 ms to 100 ms in the box-car-shaped matched filter. 3364 candidates with a S/N threshold larger than 7 were plotted and visually inspected. \par

In the post processing phase, we used the software package \textsc{DMPhase} to further refine the DM. The \textsc{DMPhase} use the Fourier-domain method, where DM is found by maximising the time derivative of normalized "intensity". To measure the intensity, the polarisation calibration is then performed with software package \textsc{PSRCHIVE}\cite{HvSM04}. We adopted the single axis model in polarisation calibration, where the differential gain and phase between the two polarisation channels are calibrated with the  injected noise signal. To reduce the dynamic spectra to a manageable size, we integrate over time and frequency to reduce the resolution. The frequency and time resolutions of the final dynamic spectra are $\approx$ 1.0~MHz and $\approx$ 0.2 ms, respectively. We store the data in the \textsc{PSRFITs}\cite{HvSM04} format, which is widely used in the community of pulsar astronomy.

\section{Data format and contents of the library}


The \textsc{psrfits} format is based on the Flexible Image Transport System (FITS){(\url{https://fits.gsfc.nasa.gov/})}\cite{fits2001}. According to FITS standards, a \textsc{psrfits} file consists of a primary header-data unit (HDU) followed by a series of extension HDUs\cite{HvSM04}. As for our data, the primary HDU contains basic information such as telescope name and its location, source location, observation time and etc. Four extension HDUs, which are in a binary table format, contain specific information related to the observation: processing history, pulsar ephemeris, tempo2 predictor and the pulsar data. Notice that there are several \textsc{psrfits} files contain more than one burst because the interval between their TOAs (time of arrivals) is quiet small. \par

\begin{figure}
    \centering
    \includegraphics{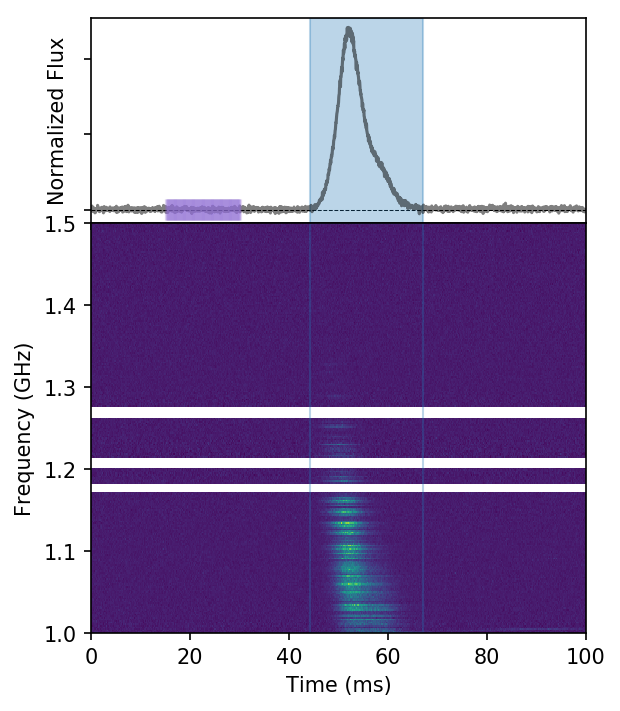}
    \caption{Example of pulse profile (upper panel) and dynamic spectra (lower panel). Purple box and blue area in the pulse profile show the baseline and pulse on-phase definition. White strips in the dynamical spectra indicate the removed channels due to the RFI.
    \label{fig:dyn}}
\end{figure}

We associate each pulse with a mask file. The mask file, formatted in plain ascii file, contains two rows of data. The first row consists two integer numbers corresponding to the boundary of pulse on-phase in the profile. The second row of mask file shows where the baseline lies in. \FIG{fig:dyn} shows the dynamic spectrum of pulse No.12 as \textcolor{black}{an example}.

\section{Statistics of data properties}

\begin{figure}
    \centering
    \includegraphics[width=1.0\textwidth]{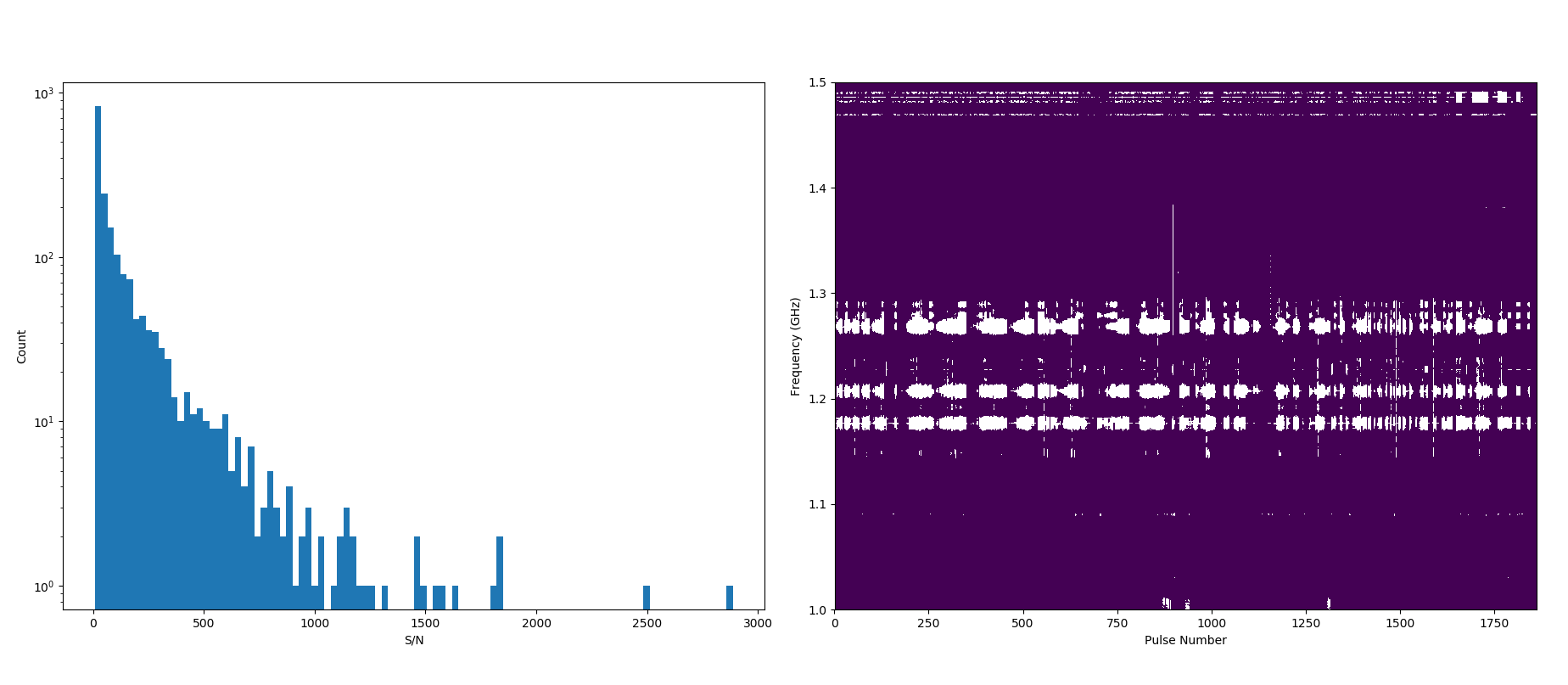}
    \caption{Left: The $S/N$ distribution of 1863 pulses. Right: The distribution of RFI zapping for all data.
    \label{fig:distri}}
\end{figure}

Our detection threshold was a signal-to-noise ratio $S/N>7$, and 1103 bright bursts reached $S/N>30$ among a total of 1863 detected bursts. The left panel of \FIG{fig:distri} shows the $S/N$ distribution of all detected bursts. The right panel of \FIG{fig:distri} shows the distribution of removed channel in frequency band. Usually, a few percent frequency channels had been removed  due to the RFI. \par

The sample completeness was determined with the following method. We simulated 10,000 mock bursts with Gaussian profile and bandpass matching the detected distributions. We then randomly injected the mock bursts into the original FAST data when no FRB was detected. The mock burst injected data are then fed to our burst-searching pipeline to compute the detection rate. The procedure shows that the fluence threshold achieving the 95\% detection probability with $S/N \ge 7$ is 53\,mJy\,ms\cite{xuheng2021}. \par

\textcolor{black}{Parameters of each burst (including burst MJD, $S/N$, DM, etc) are available in the section \textbf{Data availability} of Ref\cite{xuheng2021}.}

\section{Conclusion}

In this work, we present \textcolor{black}{a collection of dynamic spectra for 1863 FAST-detected radio bursts of FRB~20201124A} during April to June in 2021. This is the largest burst sample detected in any FRB so far. \par

The signal of FRB 20201124A is highly polarised\cite{xuheng2021}. Our dynamic spectra is  polarisation calibrated. Previous study shows that 0.5\% polarisation fidelity can be achieved with the current calibration method\cite{Luo2020Natur}. \par

The current data set is of high S/N, where 5\%, 30\% and 67\% data had S/N$\ge$ 560.63, 116.02, and 23.85, respectively. Simulation is used to determine the completeness of burst detection, where 95\% completeness fluence threshold is 53 mJy\,ms. \par

For each burst, we provide one \textsc{PSRFITs} file and one mask file. We provide the total intensity data in \textsc{PSRFITs} format, and \textcolor{black}{mask} file in ascii format which labels the burst.

\section{Data availability and related softwares}
The data that support the findings of this study are openly available in Science Data Bank at \url{https://www.doi.org/10.57760/sciencedb.j00113.00076}. The related software can be found the corresponding repositories. \par

\textsc{TransientX}: {\url{https://github.com/ypmen/TransientX}} \par

\textsc{DMPhase}: \url{https://www.github.com/DanieleMichilli/DM_phase} \par

\textsc{psrchive}: \url{http://psrchive.sourceforge.net}

\section*{Acknowledgments}
This work made use of data from the FAST. FAST is a Chinese
national megascience facility, built and operated by the National
Astronomical Observatories, Chinese Academy of Sciences.  We acknowledge the use of public data from the Fermi Science Support Center (FSSC). This work is supported by National SKA Program of China (2020SKA0120100, 2020SKA0120200), Natural Science Foundation of China (12041304, 11873067, 11988101, 12041303, 11725313, 11725314, 11833003, 12003028, 12041306, 12103089, U2031209, U2038105, U1831207), National Program on Key Research and Development Project (2019YFA0405100, 2017YFA0402602, 2018YFA0404204, 2016YFA0400801), Key Research Program of the CAS (QYZDJ-SSW-SLH021), Natural Science Foundation of Jiangsu Province (BK20211000), Cultivation Project for FAST Scientific Payoff and Research Achievement of CAMS-CAS, the Strategic Priority Research Program on Space Science, the Western Light Youth Project of Chinese Academy of Sciences, the Chinese Academy of Sciences (grants XDA15360000, XDA15052700, XDB23040400), funding from the  Max-Planck Partner Group, the science research grants from the China Manned Space Project (CMS-CSST-2021-B11,NO. CMS-CSST-2021-A11), and PKU development grant 7101502590. KJL acknowledge support from the XPLORER PRIZE. BBZ is supported by Fundamental Research Funds for the Central Universities (14380046), and the Program for Innovative Talents, Entrepreneur in Jiangsu. 
\newpage
\bibliographystyle{naturemag}

\end{document}